\documentclass[prl,twocolumn,floatfix,showpacs ]{revtex4-1}
\UseRawInputEncoding
\usepackage{amsmath}
\usepackage{amssymb}
\usepackage{graphicx}
\usepackage{dcolumn}
\usepackage{bm}
\usepackage[colorlinks=true,linkcolor=blue,anchorcolor=blue, citecolor=cyan,urlcolor=cyan]{hyperref}
\usepackage[mathlines]{lineno}
\usepackage{ulem}
\usepackage{epstopdf}
\begin{document}
\title{Hidden fully-compensated ferrimagnetism}
\author{San-Dong Guo}
\email{sandongyuwang@163.com}
\affiliation{School of Electronic Engineering, Xi'an University of Posts and Telecommunications, Xi'an 710121, China}
\begin{abstract}
Incorporating zero-net-magnetization magnets that exhibit spin-splitting into spintronics delivers key advantages: faster switching dynamics, greater immunity to destabilizing fields, lower power consumption, and markedly improved overall efficiency.  The collinear  magnets with net-zero magnetization and spin-splitting mainly include altermagnet and fully-compensated ferrimagnet,  which provide possibility to achieve  hidden spin polarization (HSP) with  net-zero spin polarization
in total but non-zero local spin polarization. In addition to proposal of hidden altermagnetism, we hereby introduce this concept of hidden fully-compensated ferrimagnetism,  where the  total spin polarization is zero, but either of the two inversion-partner sectors possesses fully-compensated ferrimagnetism with non-zero local spin polarization in the real space.
By the first-principle
calculations, we predict that $PT$-bilayer $\mathrm{CrMoC_2S_6}$  is a
possible hidden fully-compensated ferrimagnet,  showing  fully-compensated ferrimagnetic HSP, which can be separated and observed by an  out-of-plane external electric field.  Our works  provide a class of hidden spin-polarized materials that facilitates the advancement of spintronics.

\end{abstract}
\maketitle
\textcolor[rgb]{0.00,0.00,1.00}{\textbf{Introduction.---}}
Hidden spin polarization (HSP) has been proposed in centrosymmetric  nonmagnetic systems, where  the global crystal symmetry forces spin degeneracy, but the noncentrosymmetric  individual sector can   produce   visible spin-splitting effects in the real space\cite{h3}.
With the help of spin-orbit coupling (SOC),   the Dresselhaus-type   and  Rashba-type \cite{h1,h2}  momentum-dependent spin-splitting can exist in noncentrosymmetric  individual sector.  Subsequently,  a number of
layered bulk materials have been  theoretically predicted and  experimentally observed, which can   exhibit HSP\cite{h4,h5,h6,h7,h8,h9,h10}.
The HSP has also been observed in   two-dimensional (2D) $\mathrm{PtSe_2}$ by the measurement of spin- and angle-resolved
photoemission spectroscopic\cite{h11}.  More hidden physical effects have also been proposed,  for example  hidden orbital polarization and hidden Berry curvature\cite{h12,h121,h122}.
In addition to Dresselhaus-type   and  Rashba-type, the spin-splitting can also exist in magnetic materials without the assistance of SOC, such as ferromagnets, ferrimagnets and altermagnets\cite{k4,k5,k511,k512,k513}, which provide the possibility to construct individual sector of HSP system.
Here,  we focus on the collinear   magnets with net-zero magnetization, because they possess advantages of  faster switching
dynamics, greater insensitivity to destabilizing fields,  reduced power consumption and so on\cite{f0}.

The altermagnet and fully-compensated ferrimagnet of  collinear   magnets  encompass not only net-zero magnetization but also  spin-splitting.
Unlike conventional antiferromagnetism, the two sublattices of  altermagnetism  are connected by  rotational/mirror transformation
rather than by translation or inversion, which can exhibit alternating spin-splitting of  $d$-, $g$-, or $i$-wave symmetry  in Brillouin zone (BZ)\cite{k4}.
Various altermagnetic materials have been revealed  both experimentally and theoretically\cite{h13}, and even multifunctional  altermagnets have been proposed, like
valley-polarized and antiferroelectric altermagnets\cite{k8,k80,k9,k10,k10-1,k7-3-2}.
 The fully-compensated ferrimagnet represents a unique class of ferrimagnetic materials characterized by  net-zero magnetization\cite{f1,f2,f3}.
Recently,   the importance of  2D fully-compensated ferrimagnet is also emphasized\cite{f4}, which broadens  low-dimensional spintronic materials.
The two sublattices of   fully-compensated ferrimagnetism  are related by  null symmetry, and its net-zero magnetization is due to  gap-guaranteed spin quantization in one spin channel\cite{f4}.   The spin-splitting of  fully-compensated ferrimagnetism is because  the magnetic atoms with opposite spin polarization locate in the different environment, which can also be used to explain the spin-slitting  of  altermagnetism (the different environment is produced  by the orientation of surrounding atoms arrange.)\cite{f5}.  Unlike conventional antiferromagnet,  both altermagnet and fully-compensated ferrimagnet with net-zero magnetization  can produce   anomalous Halll/Nernst effect  and magneto-optical Kerr effect\cite{h13,f4}.

Recently, hidden altermagnetism has been proposed, and the system possesses  $PT$ symmetry (the joint symmetry of space inversion symmetry ($P$) and time-reversal symmetry ($T$)) with total  net-zero spin polarization,  but  either of the two inversion-partner sectors possesses altermagnetism with local non-zero  spin polarization in the real space\cite{f6}.  In this work, we extend this idea and further propose  the concept of hidden fully-compensated ferrimagnetism: the system with $PT$ symmetry  consists of two separate fully-compensated ferrimagnetic inversion partners, and it possesses   net-zero spin polarization in total, but either of the two inversion-partner sectors has local non-zero spin polarization in the real space. Although symmetry analysis suggests that antiferromagnets
hosting HSP can be classified into six types,  the ref.\cite{nc}  does not explicitly state that individual sector can be a fully-compensated ferrimagnet (From a symmetry perspective, ferrimagnetism is often subsumed under ferromagnetism.).

By first-principles calculations,  $PT$-symmetric bilayer $\mathrm{CrMoC_2S_6}$ is used  as a example to elaborate hidden fully-compensated ferrimagnetism, and an out-of-plane external electric field is applied to separate local fully-compensated ferrimagnetic  spin-splitting.
In fact, $PT$-symmetric bilayer can form a large class of hidden fully-compensated ferrimagnet, as long as fully-compensated ferrimagnet is used as the building block.
Both hidden fully-compensated ferrimagnetism and  hidden altermagnetism can provide alternate ideas to explore stable low-power spintronic devices.
\begin{figure}[t]
    \centering
    \includegraphics[width=0.40\textwidth]{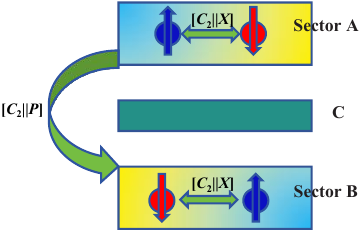}
    \caption{(Color online) For $PT$-antiferromagnet, the system consists of sector A, sector B and C, and the C plane contains the inversion center, which separates the
 unit cell into  sector A and B. The sector A and B with net-zero magnetization  are connected by the [$C_2$$\parallel$$P$].
 The spin-up and spin-down magnetic atoms of sector A and B   are connected
by [$C_2$$\parallel$$X$].  If $X$ denotes null   symmetry operation,  the sector A and B  are  fully-compensated ferrimagnets, and the system is referred to as hidden fully-compensated ferrimagnetism.  If $X$ stands for rotation or mirror symmetry operation,  the sector A and B  are  altermagnets, and  the system is called hidden  altermagnetism.}\label{sy}
\end{figure}

\textcolor[rgb]{0.00,0.00,1.00}{\textbf{Concept of hidden fully-compensated ferrimagnetism.---}}
For   $PT$-antiferromagnet, the  global spin degeneracy or no spin-splitting is constrained by symmetry:$E_{\uparrow}(k)$=$PT$$E_{\uparrow}(k)$=$P$$E_{\downarrow}(-k)$=$E_{\downarrow}(k)$.  Hoever,  the local spin polarization can be achieved by   introducing the degree of freedom of the  'layer' in the real space.  In this study, we propose the concept of \textit{Hidden fully-compensated ferrimagnetism} (\autoref{sy}).   The system consists of sector A,  B and C, and the C plane contains the inversion center,  separating the
 unit cell into  sector A and B. The sector A and B  possess  net-zero magnetization, and they  are connected by the [$C_2$$\parallel$$P$] (The $C_2$ is the two-fold rotation perpendicular to the spin axis in the spin space.), producing a $PT$-antiferromagnet.
 The spin-up and spin-down magnetic atoms of sector A and B   are connected
by [$C_2$$\parallel$$X$]  (The$X$ means null symmetry operation in the lattice space). These   means that A and B are fully-compensated ferrimagnet. If $X$ stands for rotation or mirror symmetry operation,  the sector A and B  are  altermagnets, and  the system is called hidden  altermagnetism\cite{f6}.

\begin{figure}[t]
    \centering
    \includegraphics[width=0.45\textwidth]{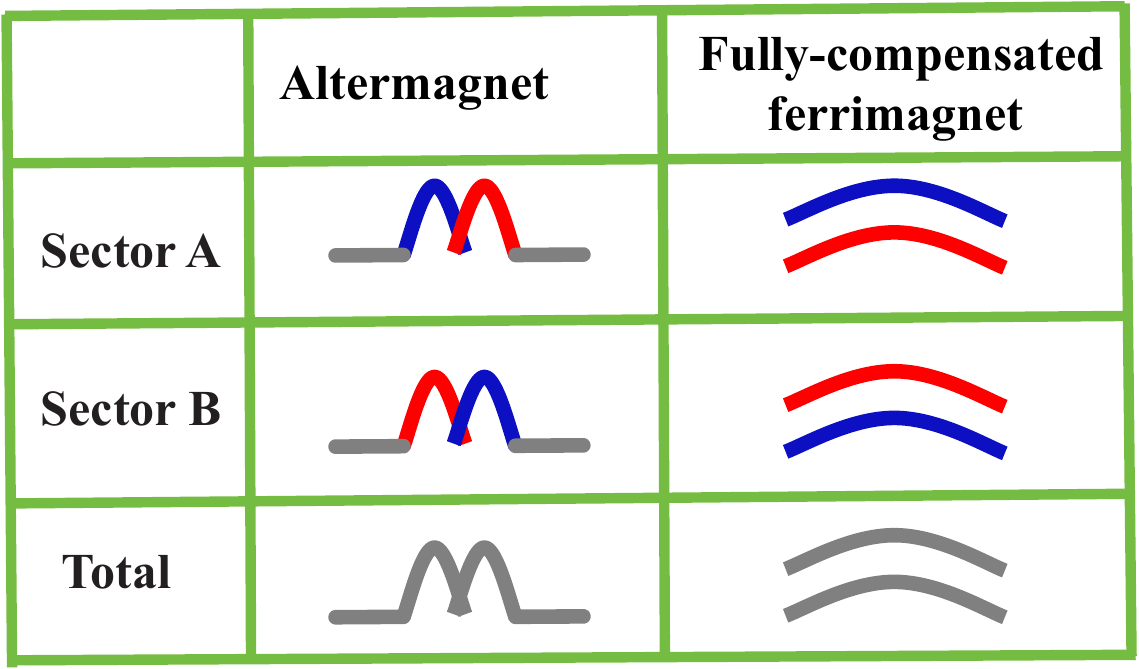}
    \caption{(Color online) The sector A and B  are  altermagnets (fully-compensated ferrimagnets),  and they   show  inverse momentum-dependent (global) spin-splitting,  giving rise to at least two-fold degenerate energy bands of $PT$-antiferromagnet.   The spin-up
and spin-down channels are depicted in blue and red, and the  gray means spin degeneracy.}\label{sy-1}
\end{figure}

The difference between hidden  altermagnetism and hidden fully-compensated ferrimagnetism is shown in \autoref{sy-1}.   For  hidden  altermagnetism , the sector A and B  are  altermagnets,   giving rise to inverse momentum-dependent  spin-splitting,   such  as $d$-wave, $g$-wave  and $i$-wave symmetry\cite{k4}. For hidden fully-compensated ferrimagnetism, the sector A and B  are composed of fully-compensated ferrimagnet,  producing inverse global spin-splitting of $s$-wave symmetry.  For both cases, the energy  bands of the whole system is at least doubly degenerate.

Searching for realistic hidden fully-compensated ferrimagnet is difficult, but bilayer stacking engineering provides a general way to achieve hidden fully-compensated ferrimagnetism.  By a similar way of building hidden  altermagnet\cite{f6}, we can construct hidden fully-compensated ferrimagnet by taking fully-compensated ferrimagnetic monolayer as the basic building unit.  The fully-compensated ferrimagnetic monolayer can be obtained from   $PT$-antiferromagnet and altermagnet by removing the link symmetry between magnetic atoms with opposite spin polarization using  external electric field, built-in electric field and isovalent alloying\cite{f4,f5,f7,f7-1,f8,f9,f10}, such as fully-compensated ferrimagnetic $\mathrm{CrMoC_2S_6}$, $\mathrm{Cr_2CHCl}$, $\mathrm{Fe_2CFCl}$, $\mathrm{Mn_2ClI}$,  $\mathrm{V_2F_7Cl}$  and so on.
In the bilayer system,  a perpendicular electric field $E$ can make the  fully-compensated ferrimagnetism
localized on each layer to be observable, producing the layer-locked  anomalous Halll/Nernst effect  and magneto-optical Kerr effect.
The spin/layer order can be reversed by reversing the direction of the electric field.
Here, we take $\mathrm{CrMoC_2S_6}$ monolayer as the basic building unit to illustrate the concept of hidden fully-compensated ferrimagnetism.

\begin{figure}[t]
    \centering
    \includegraphics[width=0.48\textwidth]{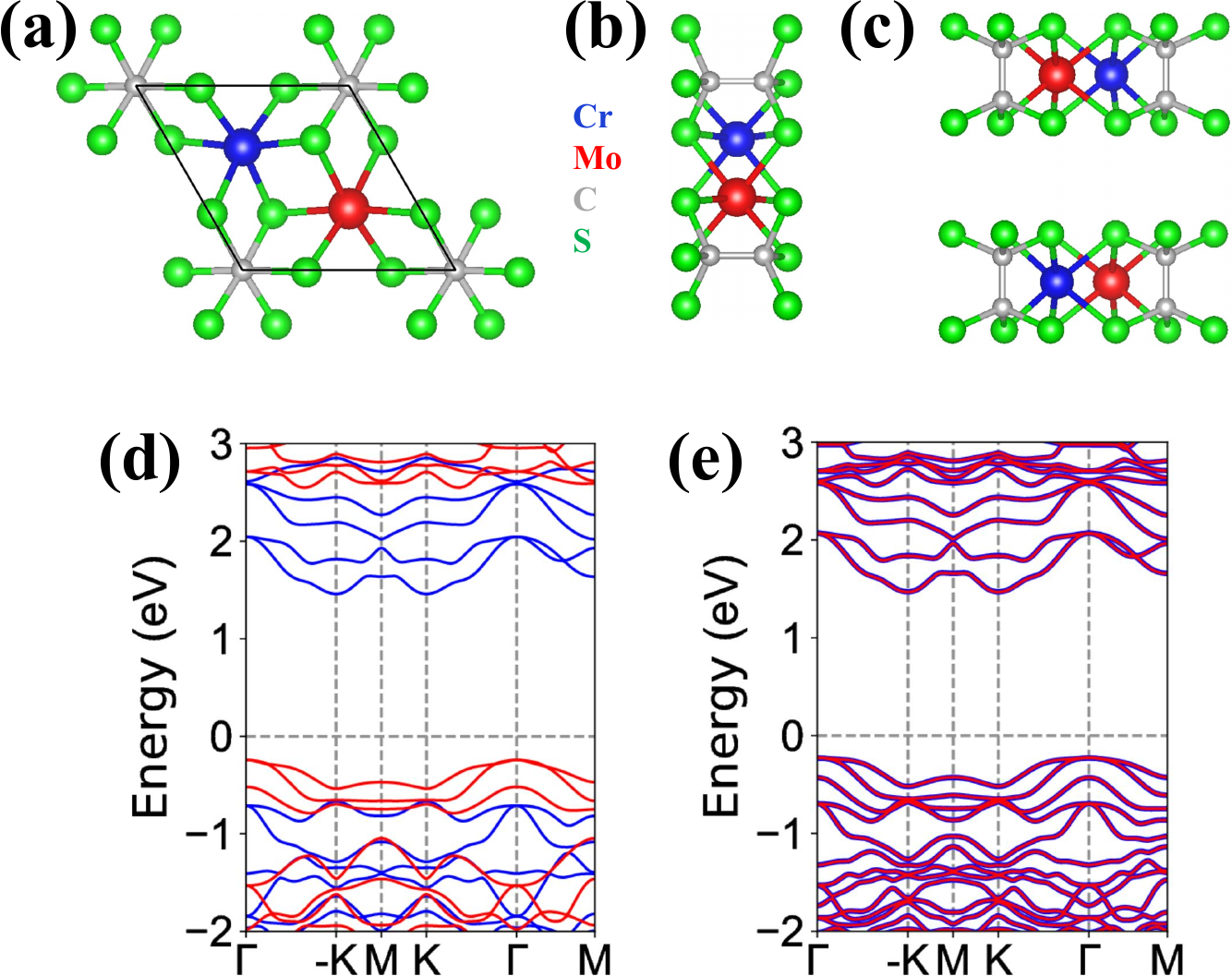}
     \caption{(Color online) The top (a) and side (b) views of  the crystal structure and  energy band structures (d) of  monolayer   $\mathrm{CrMoC_2S_6}$;  The  side (c) view of  the crystal structure and  energy band structures (e) of  $PT$-bilayer  $\mathrm{CrMoC_2S_6}$. In (a, b, c),  the blue, red, gray and green balls  represent Cr, Mo, C and S atoms, respectively. In  (d, e), the spin-up
and spin-down channels are depicted in blue and red. }\label{band}
\end{figure}

\textcolor[rgb]{0.00,0.00,1.00}{\textbf{Computational detail.---}}
We perform  spin-polarized  first-principles calculations   within density functional theory (DFT) \cite{1},  as implemented in Vienna ab initio simulation package (VASP)\cite{pv1,pv2,pv3} by using the projector augmented-wave (PAW) method. The generalized gradient approximation (GGA) of  Perdew, Burke, and Ernzerhof (PBE)\cite{pbe} as the exchange-correlation functional  is adopted. The kinetic energy cutoff  of 500 eV,  total energy  convergence criterion of  $10^{-7}$ eV, and  force convergence criterion of less than 0.001 $\mathrm{eV.{\AA}^{-1}}$ are set to obtain the accurate results.
We add Hubbard correction with $U$=3.00 eV\cite{f7-1} for $d$-orbitals of both Cr and Mo atoms within the
rotationally invariant approach proposed by Dudarev et al\cite{du}.
A slab model with a vacuum thickness of more than 15 $\mathrm{{\AA}}$ along $z$ direction is used to avoid interlayer interactions.
We sample BZ with a 13$\times$13$\times$1 Monkhorst-Pack $k$-point meshes  for structure relaxation and electronic structure calculations. The dispersion-corrected DFT-D3 method\cite{dft3} is adopted to describe the van der Waals (vdW)
interactions.  We  determine the magnetic orientation  by calculating  magnetic anisotropy energy (MAE): $E_{MAE}=E^{||}_{SOC}-E^{\perp}_{SOC}$, where $||$ and $\perp$  mean that spins lie in
the plane and out-of-plane.

\begin{figure}[t]
    \centering
    \includegraphics[width=0.48\textwidth]{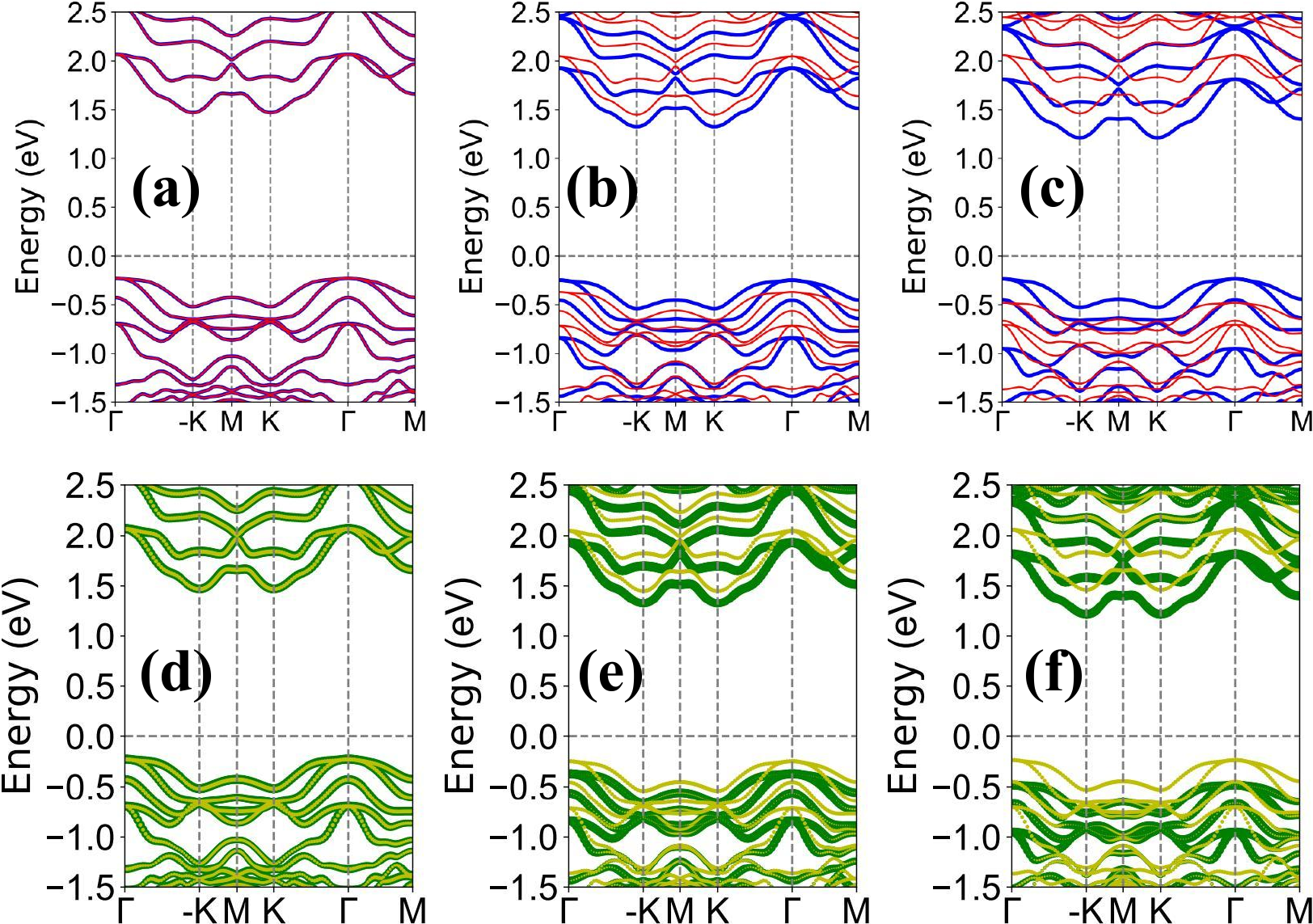}
     \caption{(Color online) For bilayer  $\mathrm{CrMoC_2S_6}$, the energy band structures (a, b, c) with layer-characteristic projection (d, e, f) at  representative $E$=+0.00,  +0.02 and +0.04  $\mathrm{V/{\AA}}$.
     In (a, b, c),  the spin-up
and spin-down channels are depicted in blue and red. In (d, e, f),
      the yellow and green represent  lower- and  upper-layer characters.}\label{band-1}
\end{figure}

\textcolor[rgb]{0.00,0.00,1.00}{\textbf{Material realization.---}}
The  $\mathrm{CrMoC_2S_6}$ monolayer has been proved to be dynamically,  mechanically and thermally stable\cite{f7,f7-1}, and its  crystal structures  are  shown in \autoref{band} (a, b),  crystallizing  in the  $P312$ space group (No.149) without  spatial inversion symmetry.  The magnetic ground state of  $\mathrm{CrMoC_2S_6}$  shows that Cr and Mo have opposite spin polarization in one primitive cell with total net-zero magnetization, and they cannot be symmetrically connected,  giving rise to fully-compensated ferrimagnetism. In fact,  the  fully-compensated ferrimagnetic $\mathrm{CrMoC_2S_6}$ can be obtained by substituting one Cr of  $PT$-antiferromagnetic $\mathrm{Cr_2C_2S_6}$ with Mo via isovalent alloying. The optimized  theoretical lattice constants are $a$=$b$=5.714 $\mathrm{{\AA}}$ by using  GGA+$U$ method.
The energy band structures of  $\mathrm{CrMoC_2S_6}$  are plotted in \autoref{band} (d)  without SOC,  which shows  a large spin-splitting around the
Fermi energy level due  to $d$ orbital mismatch between Cr and Mo atoms. The   $\mathrm{CrMoC_2S_6}$  shows an indirect gap  with  conduction band bottom (CBM)/valence band maximum (VBM) at K/$\Gamma$ point,  and  the $\mathrm{CrMoC_2S_6}$ is also  a bipolar ferrimagnetic semiconductor because of VBM and
CBM from different spin channels. The  total magnetic moment of $\mathrm{CrMoC_2S_6}$ per unit cell is strictly 0.00 $\mu_B$, and the  magnetic moments of Cr and Mo atoms are 2.88  $\mu_B$ and -2.34 $\mu_B$, respectively.

Next, we build  $P$-symmetric bilayer   $\mathrm{CrMoC_2S_6}$, and its crystal structures  are  shown in \autoref{band} (c),  crystallizing  in the  $P\bar{3}$ space group (No.147) with  spatial inversion symmetry. The intralayer AFM and interlayer FM  (AFM1), and intralayer AFM and interlayer  AFM  (AFM2) configurations (see FIG.S1\cite{bc}) are constructed to determine the magnetic  ground state of bilayer   $\mathrm{CrMoC_2S_6}$, and the AFM1 magnetic configuration satisfies $PT$ symmetry.
 Calculated results show that bilayer   $\mathrm{CrMoC_2S_6}$  possesses AFM1 ground state, which is 1.32 meV per unit cell lower than that of AFM2 case. This ensures that bilayer   $\mathrm{CrMoC_2S_6}$  is globally spin degenerate, producing hidden fully-compensated ferrimagnetism.
  In general,  the electron correlation of 4$d$ electrons is weak  than that of 3$d$ electrons. To ensure that bilayer   $\mathrm{CrMoC_2S_6}$ exhibits hidden fully-compensated ferrimagnetism, it should adopt the AFM1 ordering. Therefore, the different $U$(Cr, Mo) values are also considered to  calculate the energy difference between  AFM2 and AFM1 ordering.
From $U$(3, 2) to $U$(3, 3) to $U$(4, 3), the energy difference between  AFM2 and AFM1 ordering  changes from 2.47  meV to  1.32 meV   to  1.71 meV, and these positive values confirm the reliability of our results.
  By GGA+$U$  with AFM1 ordering,  the optimized lattice constants $a$=$b$=5.686 $\mathrm{{\AA}}$  is slightly smaller than that of monolayer. The calculated MAE is 247$\mathrm{\mu eV}$/unit cell,  implying the out-of-plane easy magnetization axis of bilayer   $\mathrm{CrMoC_2S_6}$.
 The energy band structures   are plotted in \autoref{band} (e)  without SOC,  which shows an indirect bandgap semiconductor with VBM/CBM  at K/$\Gamma$ point.
 Calculated results show that  every band is doubly degenerate due to $PT$ symmetry.

\begin{figure}[t]
    \centering
    \includegraphics[width=0.40\textwidth]{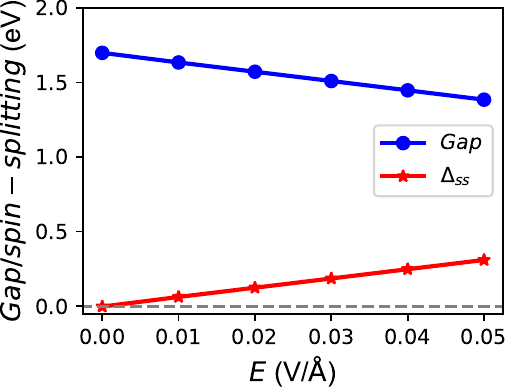}
     \caption{(Color online) For  bilayer  $\mathrm{CrMoC_2S_6}$, the energy band gap ($Gap$) and the spin-splitting ($\Delta_{ss}$) of  CBM  (The energy difference  between the first and second conduction bands at -K/K point) as a function of electric field $E$.}\label{gap}
\end{figure}

 Electric field is an effective way to break $P$ symmetry of $PT$-antiferromagnet, and then can lift the spin degeneracy\cite{f9}.  An external electric field along the $z$-direction is applied to  bilayer   $\mathrm{CrMoC_2S_6}$, and  the magnetic ground state is determined under $+z$ electric field  ($E$=0.00-0.05 $\mathrm{V/{\AA}}$)  by
the energy difference between  AFM2 and AFM1 ordering. Based on FIG.S2\cite{bc}, the   AFM1 ordering is always ground state within considered $E$ range.
The MAE  as a function of $E$ is shown in FIG.S3\cite{bc}, and  the positive  MAE confirms that the  easy axis of  bilayer   $\mathrm{CrMoC_2S_6}$  is always out-of-plane within considered $E$ range.

 Without including SOC, the energy band structures with spin- and  layer-characteristic projection  at  representative $E$=+0.00,  +0.02 and +0.04  $\mathrm{V/{\AA}}$  are plotted in \autoref{band-1}.
With applied electric field, it is clearly seen that there is  spin-splitting of $s$-wave symmetry, which is  due to layer-dependent electrostatic potential
caused by out-of-plane  electric field.   The layer-characteristic projection along with spin-polarized character shows that each layer possesses  fully-compensated ferrimagnetic spin-splitting.  The valence and conduction bands in proximity to the Fermi level exhibit identical spin characteristics but possess distinct layer attributes.
Due to identical spin characteristics of VBM and CBM, a sufficiently strong electric field can induce a half-metallic state.
The real-space segregation of spin-splitting  make hidden fully-compensated ferrimagnetism to be observed experimentally.  When the direction of electric field is reversed, the order of  spin- or layer-splitting is also reversed (see FIG.S4\cite{bc}).
\begin{figure}[t]
    \centering
    \includegraphics[width=0.48\textwidth]{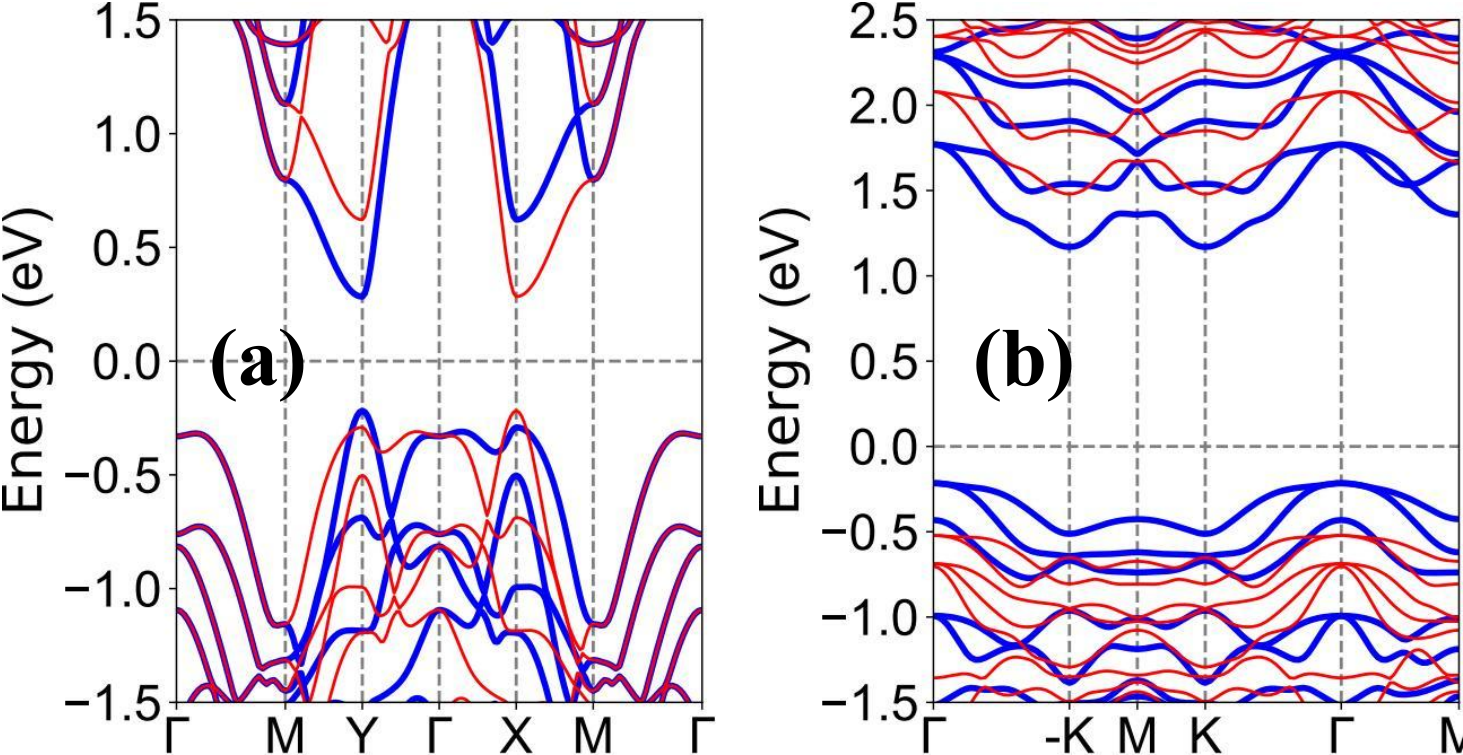}
     \caption{(Color online)  The energy band structures  of  hidden altermagnetic bilayer $\mathrm{Cr_2SO}$ (a) and hidden fully-compensated ferrimagnetic  bilayer $\mathrm{CrMoC_2S_6}$  (b)  at   $E$=+0.05  $\mathrm{V/{\AA}}$.}\label{band-2}
\end{figure}

The energy band gap  along with  the spin-splitting of CBM  as a function of electric field $E$ are plotted  in \autoref{gap}.
A linear relationship is evident between  gap/spin-splitting and electric field, and the gap decreases, while  the spin-splitting increases with increasing $E$.
The spin-splitting of CBM  can be approximately estimated by  $eEd$ with $e$ and $d$ being the electron charge and the interlayer distance of two magnetic layers\cite{k10}.
For example $E$=+0.04$\mathrm{V/{\AA}}$,  the estimated spin-splitting  is approximately 250 meV with the $d$   being  6.24 $\mathrm{{\AA}}$, which is  very close to the DFT result of 248 meV.

\textcolor[rgb]{0.00,0.00,1.00}{\textbf{Discussion and Conclusion.---}}
In experiment, both hidden altermagnetism and  hidden fully-compensated ferrimagnetism with local spin-splitting can be observed by using spin- and angle-resolved photoemission spectroscopy
(ARPES) measurements, which has been used  in measuring HSP effect of  nonmagnetic materials\cite{h7,h8,h9,h10,h11}.
The difference between hidden altermagnetism and  hidden fully-compensated ferrimagnetism lies in the symmetry of local spin-splitting. To illustrate the difference,
the energy band structures  of  hidden altermagnetic bilayer $\mathrm{Cr_2SO}$\cite{f6}and hidden fully-compensated ferrimagnetic  bilayer $\mathrm{CrMoC_2S_6}$  at   $E$=+0.05  $\mathrm{V/{\AA}}$ are plotted in \autoref{band-2}.   Bilayer $\mathrm{Cr_2SO}$ shows momentum-dependent spin-splitting with $d$-wave symmetry, while bilayer $\mathrm{CrMoC_2S_6}$  manifests global spin-splitting with $s$-wave symmetry.
\begin{figure}[t]
    \centering
    \includegraphics[width=0.40\textwidth]{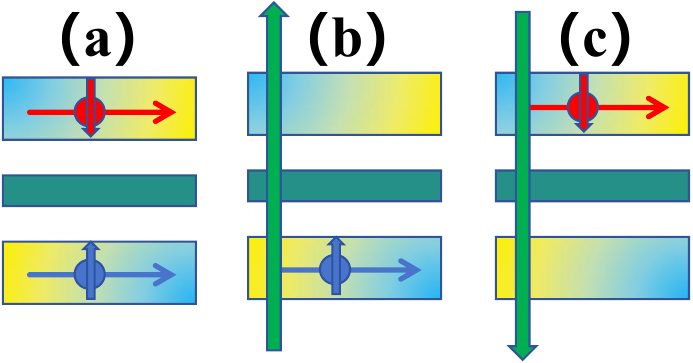}
     \caption{(Color online)  For  hidden fully-compensated ferrimagnet,  for example  bilayer $\mathrm{CrMoC_2S_6}$,  (a): without out-of-plane electric field, a hole doping can lead to  spin current in both upper and lower layers; (b): with an out-of-plane electric field, an  appropriate hole doping results in the presence of spin current only in the lower layer; (c): by reversing the direction of the electric field, the spin current is confined to the upper layer. In (b) and (c), the green arrow represents the vertical electric field.}\label{layer}
\end{figure}

 The degree of freedom of the "layer" in real space for  hidden magnetism can produce layer-locked  phenomena of  anomalous Hall/Nernst effect, nonrelativistic spin-polarized currents and the magneto-optical Kerr effect possessed  by both fully-compensated ferrimagnet and altermagnet\cite{h13,f4}. In the absence of an out-of-plane  electric field, the introduction of  carriers enables both layers to become functional (\autoref{layer} (a)). When the electric field is applied, an appropriate carrier doping can make just one layer play a role (\autoref{layer} (b)). By reversing the direction of the electric field, another layer will activate (\autoref{layer} (c)).
 Hidden altermagnet and  hidden fully-compensated ferrimagnet  constitute a special class of net-zero magnetization magnet,  which can advance the development of spintronic devices with high immunity to magnetic field disturbance.

 In summary, we propose the concept of  hidden fully-compensated ferrimagnetism with the global  spin degeneracy and the local   spin-splitting of $s$-wave symmetry, namely that the  individual sectors are fully-compensated ferrimagnets with net-zero magnetization.
By the first-principles calculations, we demonstrate  hidden fully-compensated ferrimagnetism by an extensive study of $PT$-symmetric bilayer $\mathrm{CrMoC_2S_6}$, and a vertical electric field can be used to separate and detect the  local  spin-splitting.
Our findings  provide future directions for HSP research in   magnets  with global and local  net-zero magnetization, and  motivate more theoretical and experimental works to explore other relevant physical effects.

\begin{acknowledgments}
This work is supported by Natural Science Basis Research Plan in Shaanxi Province of China   (2025JC-YBMS-008). We are grateful to Shanxi Supercomputing Center of China, and the calculations were performed on TianHe-2. We thank Prof. Guangzhao Wang for providing VASP software and helpful discussions.
\end{acknowledgments}

\end{document}